\documentclass[aps,twocolumn]{revtex4}
\usepackage{amsmath}
\usepackage{graphicx}
\begin{document}

\title{Construction and test of a moving boundary model for negative streamer discharges}
\author{Fabian Brau$^1$, Alejandro Luque$^1$, Bernard Meulenbroek$^{1,2}$, Ute Ebert$^{1,3}$ and Lothar Sch\"afer$^4$}
\affiliation{$^1$~CWI, P.O. Box 94079, 1090 GB Amsterdam, The Netherlands} \affiliation{$^2$~Faculty Electr.\
Eng., Math.\ and Comp.\ Sci., Delft Univ. Techn., The Netherlands} \affiliation{$^3$~Department of Physics,
Eindhoven University of Technology, The Netherlands} \affiliation{$^4$~Department of Physics, Universit\"at
Essen-Duisburg, Germany}

\date{\today}

\begin{abstract}
Starting from the minimal model for the electrically interacting particle densities in negative streamer
discharges, we derive a moving boundary approximation for the ionization fronts. The boundary condition on
the moving front is found to be of 'kinetic undercooling' type. The boundary approximation, whose first
results have been published in [Meulenbroek {\it et al.}, PRL {\bf 95}, 195004 (2005)], is then tested
against 2-dimensional simulations of the density model. The results suggest that our moving boundary
approximation adequately represents the essential dynamics of negative streamer fronts.
\end{abstract}

\newcommand \be{\begin{equation}}
\newcommand \ee{\end{equation}}
\newcommand \ba{\begin{eqnarray}}
\newcommand \ea{\end{eqnarray}}
\def\nn{\nonumber}
\def\np{\newpage}

\maketitle

\section{Introduction}
\label{sec1}

\begin{figure}[!hbtp]
\centerline{\includegraphics[width=6cm]{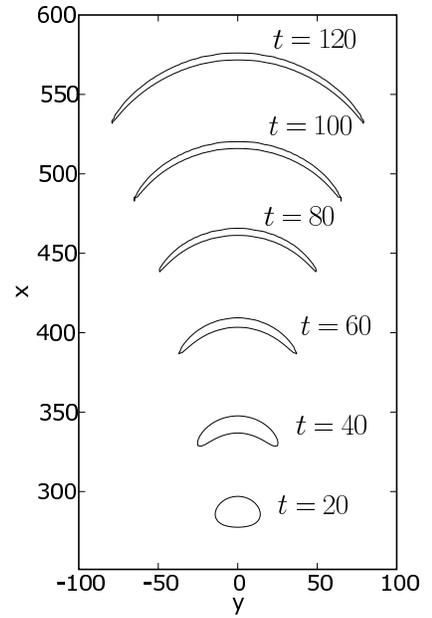}} \caption{Stages of evolution of the space charge layer
surrounding a streamer in a homogeneous background electric field of $E_0=1$ pointing downwards. Plotted is
the half maximum line of the negative space charge density $\sigma-\rho$ at each instance of time. Similar
plots in 3 spatial dimensions with radial symmetry were presented and discussed in \cite{CaroAva,CaroJCP},
the present simulation is purely 2 dimensional.} \label{fig01}
\end{figure}

Streamers are growing plasma channels extending in strong electric fields through large volumes of matter;
they determine the initial stages of electric breakdown equally in technical and natural processes
\cite{Raether,Loeb,Raizer,PSST}. Negative (anode-directed) streamers, which are the subject of this work, can
be described on a mesoscopic level by a system of reaction advection diffusion equations for electron and ion
densities coupled to the electric field. Numerical
solutions~\cite{Kunhardt,dhal85,dhal87,vite94,PRLMan,AndreaRapid,CaroAva,CaroRapid,CaroJCP} of this minimal
model reveal that the evolution of the streamer channel is dominated by a space charge layer forming around
the tip. This layer enhances the electric field in front of the streamer, which leads to rapid growth through
an efficient impact ionization by field accelerated electrons. It furthermore screens the electric field from
the streamer bulk. Fig.~\ref{fig01} illustrates the development of this space charge layer, starting from a
smooth initial ionization seed. More detailed illustrations of fully developed streamer fronts will be
presented in Sec.~\ref{curved}. In the fully developed streamer, the width of the space charge layer is much
smaller than the radius of the streamer head; this separation of scales is actually necessary for the strong
field enhancement ahead of the streamer and the field screening from the ionized interior. It suggests a
moving boundary approximation for the ionization front which brings the problem into the form of a Laplacian
interfacial growth model. Such a model was first formulated by Lozansky and Firsov~\cite{loza73} and the
concept was further detailed in \cite{PRLUWC,PREUWC,PRLMan}; solutions of such a moving boundary
approximation were discussed in \cite{Bern1,PRL05,SIAM06}.

In the moving boundary approximation presented in \cite{loza73,Bern1,PRL05,SIAM06} the space charge layer is
replaced by an infinitesimally thin interface. It is assumed that the streamer moves into a non-ionized and
electrically neutral medium, so that outside the steamer the electric potential $\varphi$ obeys the Laplace
equation
\begin{eqnarray}\label{GLvogel}
\Delta \varphi = 0   \qquad \quad \text{outside the streamer}.
\end{eqnarray}
The ionized body of the streamer is modeled as ideally conducting
\begin{eqnarray} \label {GLhund}
\varphi =\text{const}\ \qquad \text{inside the streamer}.
\end{eqnarray}
We immediately note that this latter assumption will not be essential for our analysis. The interface
separating these two regions moves with a velocity $v_n$ that depends on the local electric field
\begin{equation} \label{GLpferd}
v_{n} = v^*\Big(\nabla\varphi\Big)^+.
\end{equation}
Here and below, superscripts $^\pm$ indicate the limit value on the interface where the interface is
approached from the outside or the inside, respectively.

In earlier work \cite{loza73,Bern1}, the electric potential was taken as continuous on the streamer boundary,
$\varphi^+=\varphi^-$, which for an ideally conducting streamer interior implies that the moving interface is
equipotential, $\varphi^+=\text{const}$. However, due to the small but finite width of the physical space
charge layer this assumption is unfounded. Rather in moving boundary approximation $\varphi$ must be
discontinuous at the interface. In~\cite{PRL05,SIAM06} we modeled the potential difference across the
interface as
\begin{equation}\label{GLkatze}
\varphi^+-\varphi^- = \ell\, \hat{{\bf n}} \cdot (\nabla \varphi)^+,
\end{equation}
where the length parameter $\ell$ accounts for the thickness of the ionization front and $\hat{{\bf n}}$ is
the exterior normal on the interface. Derivation and discussion of this boundary condition is the subject of
the present paper. In the context of crystal growth from undercooled melts such a boundary condition is known
as kinetic undercooling.

Clearly the model (\ref{GLvogel})--(\ref{GLkatze}) is intimately related to moving boundary models for a
variety of different physical phenomena like viscous fingering (see, e.g., \cite{R3,R4}) or crystal growth
(see, e.g., \cite{Saito,Pom}).
To derive the model in the context of streamer motion, our starting point is the minimal streamer model that
applies to anode-directed discharges in simple gases like nitrogen or argon. Cathode-directed discharges or
discharges in composite gases like air involve additional physical mechanisms~\cite{Raizer}.

In Sec.~\ref{minimal} of the present paper we describe briefly the minimal streamer model. If diffusion is
neglected the model allows for planar shock fronts moving with constant velocity in an externally applied,
time independent electric field, and some properties of these solutions are recalled in Sec.~\ref{planar}.
Based on these results, in Sec.~\ref{construction}, supplemented by the appendix, we present a rigorous
derivation of the boundary condition (\ref{GLkatze}), valid for planar fronts in strong electric fields. The
relation of our model to other moving boundary models is briefly discussed in Sec.~\ref{discussion}. The
crucial question whether the model also applies to curved ionization fronts in weaker external fields is
considered in Sec.~\ref{curved}. Based on numerical solutions of the minimal model in two-dimensional space
we argue that our moving boundary model indeed captures the essential physics of fully developed (negative)
streamer fronts. Our conclusions are summarized in Sec.~\ref{conclusions}.



\section{Collection of some previous results}
\label{collection}
\subsection{The minimal streamer model}
\label{minimal}

The model for negative streamers in simple non-attaching gases like nitrogen and argon as used
in~\cite{dhal87, vite94, PRLUWC, CaroRapid, CaroJCP} consists of a set of three coupled partial differential
equations for the electron density $\sigma$, the ion density $\rho$ and the electric field ${\bf E}$. In
dimensionless units, it reads
\begin{eqnarray} \label{PDE1}
\partial_t \sigma + \nabla \cdot {\bf j}_e
&=&|\sigma{\bf E}| \;\alpha(|{\bf E}|),
\\
\label{PDE2}
\partial_t \rho &=& |\sigma{\bf E}| \;\alpha(|{\bf E}|),
\\
\label{PDE3}
\nabla \cdot {\bf E}&=&\rho-\sigma, \quad {\bf E} = -\nabla \varphi.
\end{eqnarray}
The first two equations are the continuity equations for the electrons and the ions while the last is the
Coulomb equation for the electric field generated by the space charge $\rho-\sigma$ of electrons and ions.
${\bf j}_e$ is the electron particle current which we here take as the drift current only
\begin{equation}
    \label{current}
    {\bf j}_e=-\sigma{\bf E}.
\end{equation}
(For the effect of a diffusive contribution to the current, see a recent summary in section 2 of
\cite{gianne} and Sec.~\ref{curved} of the present paper.) The current of the much heavier ions is
neglected. $|\sigma{\bf E}| \;\alpha(|{\bf E}|)$ is the generation rate of additional electron ion pairs; it
is the product of the absolute value of the current $|\sigma{\bf E}|$ times the effective cross section
$\alpha(|{\bf E}|)$ which is taken as field dependent; an old and much used form for $\alpha(|{\bf E}|)$ is
the Townsend approximation
\begin{equation}
\label{alphaTown}
\alpha(|{\bf E}|)=e^{-1/|{\bf E}|},
\end{equation}
but our analysis holds for the more general case of
\begin{eqnarray}
\label{alphagen}
\alpha(0)&=&0, \quad \frac{d\alpha(|{\bf E}|)}{d{|{\bf E}|}}
\ge0, \nonumber \\
\alpha(|{\bf E}|)&\underset{|{\bf E}|\gg1}{=}&1-{\cal O}\left(\frac1 {|{\bf E}|}\right).
\end{eqnarray}

The dimensional analysis reducing the physical equations to the dimensionless model defined above can be
found in many previous papers, e.g., in~\cite{PRLUWC,PSST, PREUWC}. We only note that the intrinsic time and
length scales are defined in terms of the electron mobility and the effective ionization cross section and
thus are determined by the physics on microscopic scales.

\subsection{Planar ionization fronts}
\label{planar}

We here recall essential properties of planar negative ionization fronts as derived in \cite{PREUWC,Man04}.
We consider ionization fronts propagating in the positive $x$ direction into a medium that is completely
non-ionized beyond a certain point $x_f(t)$
\begin{eqnarray}
&\sigma=0=\rho & \quad \text{for} \quad x>x_f(t),\nn \\
&\sigma>0& \quad \text{for} \quad x<x_f(t).
\end{eqnarray}
Far ahead the front, the electric field is taken to approach a constant value:
\begin{equation}
    \label{farfield}     {\bf E} \underset{x\rightarrow +\infty}{\longrightarrow} - E_0\; {\hat{\bf
x}},~~~E_0>0,
\end{equation}
where ${\hat{\bf x}}$ is the unit vector in the $x$-direction. For a planar front, ${\bf E}$ evidently can
depend only on $x$ and $t$, and Eqs.~(\ref{PDE3}), (\ref{farfield}) yield a constant field in the non-ionized
region
\begin{equation}
    \label{fieldnonionized}
    {\bf E}=-E_0 \; {\hat{\bf x}}, \quad x\ge x_f.
\end{equation}

The planar solution of the model takes the form of a uniformly translating shock front moving with velocity
\begin{equation}
\label{velocity}
v=\partial_t x_f=-E\big|_{x=x_f}=E_0.
\end{equation}
In the comoving coordinate
\begin{equation}
\label{xi}
\xi=x-x_f(0)-vt,
\end{equation}
a discontinuity of the electron density is located at $\xi=0$, while the ion density $\rho$ and the electric
field $E$ are continuous.

\begin{figure}[!hbtp]
\includegraphics[width=8cm]{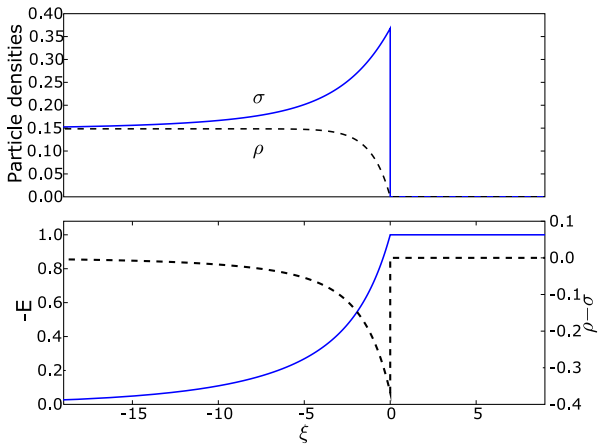}
\caption{Densities and field in a planar ionization front in a far field $E_0=1$ as a function of the comoving
coordinate $\xi$ (\ref{xi}). The front moves to the right with velocity $v=E_0$.
Upper panel: electron density $\sigma$ (solid line) and ion density $\rho$ (dashed line);
lower panel: electric field (solid line, axis on the left) and space charge density $\rho-\sigma$ (dashed line,
axis on the right). For $\alpha(|{\bf E}|)$ the Townsend approximation (\ref{alphaTown}) was used.}
\label{fig02}
\end{figure}

Fig.~\ref{fig02} that similarly has appeared in \cite{Man04}, illustrates the spatial profiles in such a
uniformly translating front for $E_0=1$. In the non-ionized region $\xi>0$, we simply have $\rho=0=\sigma$
and $E=-E_0$. In the ionized region $\xi<0$, the propagating electron front creates additional electrons and
ions as long as $\alpha(E)>0$, therefore the density of immobile ions $\rho$ increases monotonically behind
the front. The electrons move such as to screen the conducting interior from the applied electric field. They
form a layer of nonvanishing space charge $\rho-\sigma$ that suppresses the field $E$ behind the front.

Analytically the solution in the ionized region $\xi<0$ is given implicitly by the equations
\begin{eqnarray}
\label{sigma}
\sigma(E)&=&\frac{E_0}{E_0-|E|}\;\rho(E),\\
\label{rho}
\rho(E)&=&\int_{|E|}^{E_0}\alpha(\eta)\, d\eta,\\
\label{E} -\xi&=&\int_{E(\xi)}^{-E_0}\frac{E_0+\eta}{\rho(\eta)}\,\frac{d\eta}\eta.
\end{eqnarray}
In the limit $E_0\gg 1$, these equations will be further evaluated below.
We note that at the shock front, the electron density jumps from zero to
\begin{equation}
    \label{edensjump}     \sigma=E_0\, \alpha(E_0),
\end{equation}
and for $\xi\rightarrow - \infty$ it approaches the value
\begin{equation}
    \label{edensminfty}
    \sigma(-\infty)=\int_0^{E_0} \alpha(\eta)\, d\eta.
\end{equation}
Far behind the front, where ${\bf E}$ is so small that $\alpha(|{\bf E}|)\simeq 0$, the final relaxation of
{\bf E} and of $\sigma$ is exponential in space: $\sim e^{-\lambda \xi}$, where
\begin{equation}
    \label{lambdaminus}     \lambda=\frac{\sigma(-\infty)}{v}=\frac{1}{E_0}\int_0^{E_0}\alpha(\eta)\, d\eta
\stackrel{E_0\to \infty}{\longrightarrow} 1.
\end{equation}




\section{The moving boundary model}
\label{moving}

\subsection{Construction}
\label{construction}


The results reviewed in the previous section (see also Fig.~\ref{fig01}) show that a layer of space charge
$\rho-\sigma$ screens the electric field from the streamer interior. For strong applied electric field $-
E_0$, the thickness of this layer defines some small inner scale of the front, while on the large outer
scale, the streamer will be approximated as a sharp interface separating an ionized but electrically neutral
region inside the streamer from the non-ionized charge free region outside the streamer; this substantiates
the assumptions underlying the moving boundary model treated in \cite{PRL05,SIAM06}.

Being a shock front solution of Eq.~(\ref{PDE1}), the interface always moves with normal velocity
\begin{equation} \label{GLmaus}
v_n  = \hat{{\bf n}} \cdot (\nabla\varphi)^+,
\end{equation}
where $\hat{{\bf n}}$ is the unit vector normal to the interface pointing into the exterior region; this
equation generalizes Eq.~(\ref{velocity}). We recall that $^+$ indicates that the expression is evaluated by
approaching the interface from outside the streamer.

As mentioned in the introduction, in the context of electric breakdown the moving boundary model
(\ref{GLvogel}), (\ref{GLhund}), (\ref{GLmaus}) has been formulated some time ago by Lozansky and
Firsov~\cite{loza73}. To complete the model, a boundary condition at the interface is needed, and in
Ref.~\cite{loza73} continuity of the potential at the interface was postulated:
\begin{equation} \label{GLeber}
\varphi^+ = \varphi^-  = \text{const}.
\end{equation}
However, Fig.~\ref{fig02} clearly shows that crossing the screening layer, $\varphi(\xi)$ will increase,
which amounts to a jump $\varphi^+ > \varphi^-$ in the interface model. The size of this jump depends on the
electric field ${\bf E}^+$ at the interface and for a planar front is easily determined from Eq.~(\ref{E}).
We note that in the framework of the PDE-model, (Sec.~\ref{collection}), $\varphi^\pm$ are to be identified
as
\begin{equation}
\varphi^-=\varphi(-\infty), \quad \varphi^+=\varphi(0),
\end{equation}
\begin{equation} \label{rp2}
\varphi^+-\varphi^-=-\int_{-\infty}^0 E\, d\xi.
\end{equation}
Since $E(\xi)$ according to (\ref{rho}), (\ref{E}) is a monotonically decreasing function for $\xi<0$, we can
integrate by parts
\begin{eqnarray} \label{rp4}
\varphi^+-\varphi^-&=&-\int_{-\infty}^0 E\, d\xi=(\xi E)\Big|_{-\infty}^0
+\int_0^{E^+} \xi\, dE \nonumber \\&=&\int_0^{E^+} \xi\, dE.
\end{eqnarray}
The last identity holds since either $E$ or $\xi$ vanish on the integration boundaries. For a planar front,
we have $E^+ = - E_0$, but we here keep $E^+$ to stress the dependence on the field at the front position
$\xi = 0$.

While $E(\xi)$ is known only implicitly as $\xi=\xi(E)$ in Eq.~(\ref{E}), the partial integration now allows
us to evaluate the integral explicitly by substituting Eq.~(\ref{E}) in (\ref{rp4}):
\begin{eqnarray} \label{rp5}
\varphi^+-\varphi^-&=& -\int_0^{E^+} dE \int_{E}^{E^+}\frac{\eta-E^+}{\rho(\eta)}\;\frac{d\eta}{\eta}
\\
\label{rp6}
&=&-\int_0^{E^+} \frac{\eta-E^+}{\rho(\eta)}\, d\eta.
\end{eqnarray}
$\rho(\eta)$ is given in Eq.~(\ref{rho}). This result explicitly shows that in the interface model the
potential is discontinuous across the boundary, where the size of the discontinuity depends on the electric
field right ahead of the ionization front.

Evaluating Eq.~(\ref{rp6}), in the appendix we derive bounds showing that $\varphi^+-\varphi^-\approx-E^+ +
\text{ const}$ for large $|E^+|$. We here present a simpler argument yielding only the leading term. It is
based on direct evaluation of Eqs.~(\ref{rho}), (\ref{E}), written as
\begin{eqnarray}
\frac {\rho(E)} {|E^+|} &=&\int_{\frac{E}{E^+}}^1 \alpha \left( \eta\, |E^+| \right) d\eta
\label{GLstier} \\
\xi &=& |E^+| \int_{\frac{E(\xi)}{E^+}}^1  \frac{\nu - 1} {\rho (\nu E^+)} \frac{d\nu}{\nu}. \label{GLesel}
\end{eqnarray}
We now take the limit $|E^+| \to \infty$ in Eq.~(\ref{GLstier}), with $E/E^+ > 0$ fixed. The asymptotic
behavior (\ref{alphagen}) of $\alpha(|{\bf E}|)$ yields
\begin {equation}\label{GLsittich}
\frac {\rho(E)} {|E^+|} = 1 - \frac {E} {E^+}.
\end {equation}
Substituting this result into Eq.~(\ref{GLesel}) we find
\begin{displaymath}
\xi = - \int_{E(\xi)/{E}^+}^1 \frac {d\nu}{\nu},
\end{displaymath}
yielding a purely exponential front profile
\begin {equation}\label{GLtaube}
E(\xi)=E^+ e^{\xi}.
\end {equation}
This result means that the exponential decay of the space charge layer (\ref{lambdaminus}), that holds far
behind the front for all $|E^+| > 0$, for $|E^+|\gg1$ is actually valid throughout the complete front up to
$\xi=0$. Substituting Eq.~(\ref{GLtaube}) into Eq.~(\ref{rp2}), we find $\varphi^+ - \varphi^- = -E^+$.

The more precise argument given in the appendix shows that $ (\varphi^+ - \varphi^-)/|E^+|$ decreases
monotonically with $|E^+|$ and is bounded as
\begin {eqnarray} \label{GLadler}
0  &\leq& \frac {\varphi^+ - \varphi^-} {|{E}^+|} - 1 \\ &\leq& \frac {2\,b} { |E^+|} + {\cal O} \left(
\left({\frac{b}{|E^+|}}\right)^2 \ln \left(\frac{|E^+|} {b}\right) \right) \mbox{ for } |{\bf E}^+| \to
\infty, \nonumber
\end {eqnarray} where $b > 0$ is some constant. The result
\begin {equation} \label{GLhase}
\varphi^+-\varphi^- =- E^+ +b'+{\cal O} \left(\frac{\ln|E^+|}{E^+}\right)
\end {equation}
follows. It shows that the first correction to the leading behavior $\varphi^+  -  \varphi^- \sim - E^+$ is
just a constant, not a logarithmic term. We thus can choose  the gauge of $\varphi$ as $\varphi^-  +  b'  =
0$ to find $\varphi^+  =  - E^+  +  {\cal O}  (\ln|E^+| / E^+)$.

The simplest generalization of this result to curved fronts in strong fields, $|E^+|\gg1$, suggests the
boundary condition
\begin{equation} \label {GLkaninchen}
\varphi^+ = \hat{{\bf n}} \cdot (\nabla \varphi)^+,
\end{equation}
replacing the Lozansky-Firsov boundary condition (\ref{GLeber}). Boundary condition~(\ref{GLkaninchen})
together with the Laplace equation $\Delta\varphi=0$~(\ref{GLvogel}) and the interfacial velocity ${\bf
v}=\nabla\varphi$~(\ref{GLmaus}) define our version of the moving boundary model describing the region
outside the streamer and the consecutive motion of its boundary.

\subsection{Discussion}
\label{discussion}

The model formulated here belongs to a class of Laplacian moving boundary models describing a variety of
phenomena. In particular, it is intimately related to the extensively studied models of viscous
fingering~\cite{R3,R4} and solidification in undercooled melts~\cite{Saito,Pom}. 
In all these models the boundary separates an interior region from an exterior region, where the relevant
field obeys either the Laplace equation or a diffusion equation, and the velocity of the interface is
determined by the gradient of this field.

If we replace the boundary condition (\ref{GLkaninchen}) by (\ref{GLeber}): $\varphi^+ = \text {const.}$, the
model becomes equivalent to a simple model of viscous fingering where surface tension effects are neglected.
This ``unregularized'' model is known to exhibit unphysical cusps within finite time~\cite{R2,R1}. To
suppress these cusps, in viscous fingering a boundary condition involving the curvature of the interface is
used. The physical mechanism for this boundary condition is surface tension. As mentioned in the
introduction, the kinetic undercooling boundary condition (\ref{GLkaninchen}) is used in the context of
solidification. In that case, however, the relevant temperature field obeys the diffusion equation. From a
purely mathematical point of view, our model with specific conditions on the outer boundaries far away from
the moving interface has been analyzed in~\cite{R6,R5,R9}. It has been shown~\cite{R9} that with outer
boundary conditions appropriate for Hele-Shaw cells, the kinetic undercooling condition selects the same
stable Saffman-Taylor finger configuration as curvature regularization. Furthermore it has been
proven~\cite{R6,R5} that an initially smooth interface stays smooth for some finite time. This regularizing
property of the boundary condition~(\ref{GLkaninchen}) is also supported by our previous and ongoing
work~\cite{PRL05,SIAM06,R10}.

In applying an interface model to streamer propagation, an important difference from the other physical
systems mentioned above must be noted. For the other systems mentioned the moving boundary model directly
results from the macroscopic physics, irrespective of the motion of the boundary: a sharp interface with no
internal structure a priori is present. In contrast, a streamer emerges from a smooth seed of electron
density placed in a strong electric field, and the screening layer that is an essential ingredient of the
moving boundary model, arises dynamically, with properties determined by the electric field and thus coupled
to the velocity of the boundary. The model therefore does not cover the initial ``Townsend'' or avalanche
stage of an electric discharge~\cite{CaroAva} that is also visible in Fig.~\ref{fig01}, but can only be
applied to fully developed negative streamers. Furthermore, being explicitly derived for planar fronts, the
validity of the boundary condition for more realistic curved streamer fronts has to be tested. This issue is
discussed in the next section.


\section {Curved streamer fronts in two dimensions}
\label{curved}

\subsection{Illustration of numerical results for fully developed streamers}

We solve the PDE-model (\ref{PDE1})-(\ref{PDE3}) in two dimensions, using the numerical code described in
detail in \cite{CaroJCP}. (Previous simulation work was in three spatial dimensions assuming radial symmetry
of the streamer, the results are very similar.) In the electron current ${\bf j}_e$, besides the drift term
$-\sigma\,{\bf E}$, a diffusive contribution is taken into account:
\begin{equation}\label{GLuhu}
{\bf j}_{e} = -\sigma {\bf E}-D \nabla \sigma.
\end{equation}
This clearly is adequate physically, and on the technical level it smoothes the discontinuous shock front.
The price to be paid is some ambiguity in defining the position of the moving boundary, see below.

Planar fronts with $D > 0$ have been analyzed in \cite{PRLUWC,PREUWC,gianne}, for further discussion and
illustrations, we refer to section 2 of \cite{gianne}. It is found that diffusion enhances the front velocity
by a term
\begin{displaymath}
v_{n,D} =2 \sqrt{D|E^+|\alpha(|E^+|)},
\end{displaymath}
that has to be added to the velocity $v_{n,\text{drift}} = |E^+|$. Furthermore, diffusion creates a leading
edge of the electron density in forward direction which decreases exponentially on scale
\begin{displaymath}
\ell_D = \sqrt{\frac{D}{|E^+|\alpha(|E^+|)}}.
\end{displaymath}
This scale has to be compared to the thickness of the screening layer for $D = 0$: $\ell_\alpha\sim
1/\alpha(|{E}^+|)$. For large $|E^+|$ and small D both the ratios $v_{n,D}/v_{n,\text{drift}}$ and
$\ell_D/\ell_\alpha$ are of order $\sqrt{D/|E^+|} \ll 1$. This, by itself, does not imply that diffusion can
be neglected since the term $D\nabla\sigma$ is a singular perturbation of the diffusion free model. However,
in our numerical solutions the main effect of diffusion is found to amount to some rescaling of the
parameters in the effective moving boundary model, see below. This is consistent with the observation that
for long wave length perturbations of planar fronts, the limit $D\to0$ is smooth~\cite{gianne}.

In our numerical calculations, we take $D = 0.1$. For $\alpha(|{\bf E}|)$ the Townsend form (\ref{alphaTown})
is used. We start with an electrically neutral, Gaussian shaped ionization seed, placed in a constant
external electric field ${\bf E}=-E_0\, \hat{{\bf x}}$. We performed runs for $0.5 \leq E_0 \leq 2$. Since
the thickness of the screening layer decreases with increasing $E_0$, higher fields need better numerical
resolution, enhancing the numerical cost considerably. In view of the results shown below we do not expect
qualitative changes for $E_0>2$.

The system (\ref{PDE1})--(\ref{PDE3}) is solved numerically with a spatial discretization of finite
differences in adaptively refined grids and a second-order explicit Runge-Kutta time integration, as
described in detail in \cite{CaroJCP}, with the difference that the method described there was for three
dimensional streamers with cylindrical symmetry and here we adapted it to truly two-dimensional systems. The
highest spatial resolution in the area around the streamer head was $\Delta x = \Delta y = 1 / 4$ for all
simulations.

\begin{figure}[!hbtp]
\includegraphics[width=0.4\textwidth]{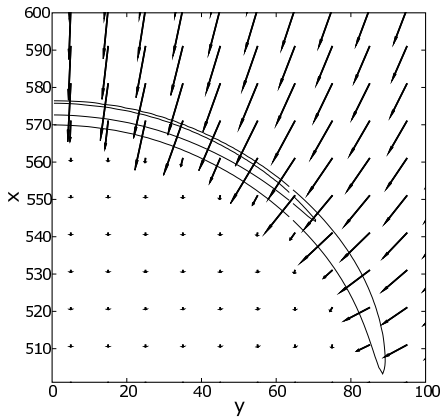}
\caption{Zoom into the right half of the symmetric streamer head at time step $t=120$ of Fig.~\ref{fig01}
in a background field of $E_0=1$. Shown are level lines of negative space charge density $\sigma-\rho$ (at
levels 1/3 and 2/3 of the maximal density) and arrows indicating the local electric field at the foot point
of the arrow.} \label{fig03}
\end{figure}

\begin{figure}[!hbtp]
\includegraphics[width=0.5\textwidth]{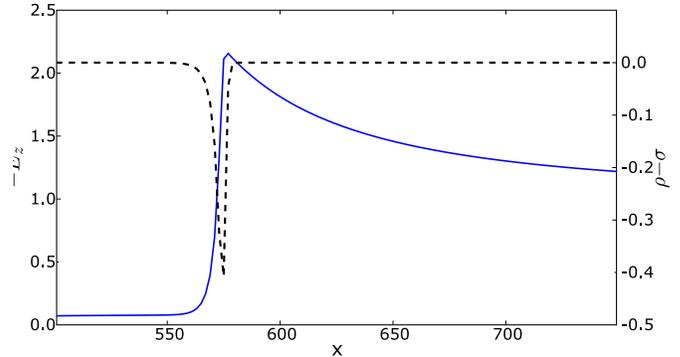}
\caption{The same case as in Fig.~\ref{fig03}, plotted is now the negative electric field (solid line, axis
on the left) and the space charge density (dashed line, axis on the right) on the axis of the streamer. The
plot can be compared to the lower panel in Fig.~\ref{fig02} showing a planar front with vanishing diffusion.}
\label{fig04}
\end{figure}

For external field $E_0=1$, Fig.~\ref{fig01} illustrates the temporal evolution of the streamer head. We see
that initially, a screening layer forms out of an ionization avalanche, this process is discussed in detail
in \cite{CaroAva}. The width of the layer rapidly reaches some almost constant value that depends on $E_0$.
The head develops into a somewhat flattened semicircle, with the radius increasing with time. This stage of
evolution will be addressed as the fully developed streamer. Fig.~\ref{fig03} zooms into the streamer head at
this stage, showing lines of constant charge density $\rho-\sigma$ together with electric field vectors.
Evidently screening
is not complete. A small, essentially constant field exists behind the streamer head. This is illustrated in
Fig.~\ref{fig04} that shows the variation of the electric field and of the excess charge along the symmetry
line $y=0$. This figure shows also that the spatial positions of the maxima of $|{\bf E}|$ and of
$\sigma-\rho$ do not coincide precisely; in fact, the maximum of $|{\bf E}|$ lies within the diffusive
leading edge of the front; the small width $\ell_D$ of this diffusive layer
replaces the jump of the electron density for $D=0$. We furthermore observe that the field behind the front
is suppressed by about a factor of $1/20$ compared to the maximal value $|{\bf E}| \approx 2.1$, or equivalently
to $\approx\!E_0/10$. This screening increases with increasing background electric field $E_0$:
from $\approx\!E_0/7$ for $E_0 = 0.5$ to $\approx\!E_0/20$ for $E_0 = 2.0$. (The maximal field enhancement
in these cases is $\gtrsim 2\;E_0$ in the fully developed streamer briefly before branching.)

\subsection{Test of the assumptions of the moving boundary model}

The moving boundary model is concerned only with the exterior region. Recalling the defining equations
\begin{eqnarray} \label{GLratte}
\Delta \varphi  &= & 0,     \qquad\quad \text{in exterior region}, \\ \label{GLochse} v_n &=& - \hat{{\bf n}}
\cdot {\bf E}^+,\\ \label{GLsau} \varphi^+ & =& - \hat{{\bf n}} \cdot {\bf E}^+,
\end{eqnarray}
we note that all explicit reference to the physics in the interior is absent, notwithstanding our derivation
in Sec.~\ref{moving}. Now the first of these equations evidently holds as soon as the diffusive leading edge
of the electron density has a negligible space charge density. Also the second relation holds for any smooth
shock front ($D = 0$) of the PDE-model. The boundary condition (\ref{GLsau}), however was derived only for
planar fronts in strong external fields $E_0\gg 1$.

\begin{figure}[!hbtp]
\includegraphics[width=0.5\textwidth]{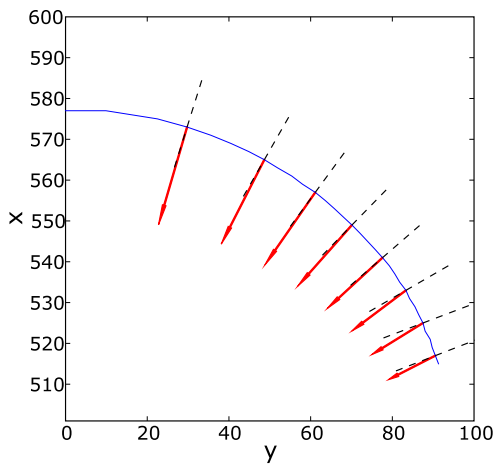}
\caption{Again the situation of Figs.~\ref{fig03} and \ref{fig04}. Shown is now the boundary defined by the
maximum of $E$ on normal intersections (solid line), the local directions of the surface normal (dashed
lines) and the local electric fields (solid arrows).} \label{fig05}
\end{figure}

To check whether the moving boundary model adequately represents the evolution of curved streamer fronts for
small diffusion and external fields of order unity, we first have to choose a precise definition of the
interface. As illustrated in Figs.~\ref{fig03} and \ref{fig04}, the screening layer is fairly thin, but
nevertheless it involves the two length scales $\ell_D$ and $\ell_\alpha$ and thus shows some intrinsic
structure. We define the moving boundary to be determined by the maximum of $|{\bf E}(x,y)|$ along
intersections perpendicular to the boundary. In precise mathematical terms a parameter representation
($x_b(u), y_b(u)$) of the boundary obeys the equation
\begin{displaymath}
0 = \hat{{\bf n}}(u) \cdot \nabla|{\bf E}||_{x =x_{b}(u), y=y_{b} (u)},
\end{displaymath}
where
\begin{displaymath}
\hat{{\bf n}}(u) \sim \left(\frac {dy_b} {du}, - \frac {dx_b}{du} \right)
\end{displaymath}
is the normal to the boundary at point $(x_b (u), y_b(u))$. To motivate this choice we note that the moving
boundary model explicitly refers only to ${\bf E}$ and not to the excess charge distribution. In practice we
determine ($x_b (u), y_b (u)$) by first searching for the maxima of $|{\bf E}|$ along horizontal or vertical
intersections, and we then iteratively refine the so determined zero order approximation. We always follow
the boundary up to the point where the local value of $| {\bf E}^+ |$ equals $E_0$. This covers all the head
of the streamer, where the essential physics takes place. Fig.~\ref{fig05} shows the resulting boundary
corresponding to the snapshot of Fig.~\ref{fig03}. We observe that the direction of ${\bf E}$ is close to,
but does not precisely coincide with the normal direction on the interface (except on the symmetry axis, of
course). The boundary is not equipotential but a small component of the electric field tangential to the
boundary drives the electrons toward the tip. This effect counteracts the stretching of the space charge
layer perpendicular to the direction of streamer motion, (see Fig.~\ref{fig01}), which in itself would lead
to a weakening of screening.

With the so defined interface we have checked that $v_n$ depends linearly on $E^+$ within the numerical
precision, therefore Eq.~(\ref{GLochse}) holds, except for an increase of the ratio $v_n/|\hat{{\bf n}}\cdot
{\bf E}^+|$, which is an expected effect of diffusion~\cite{PRLUWC,PREUWC,gianne}; this effect can be
absorbed into a rescaling of time.

\begin{figure}[!hbtp]
\includegraphics[width=0.5\textwidth]{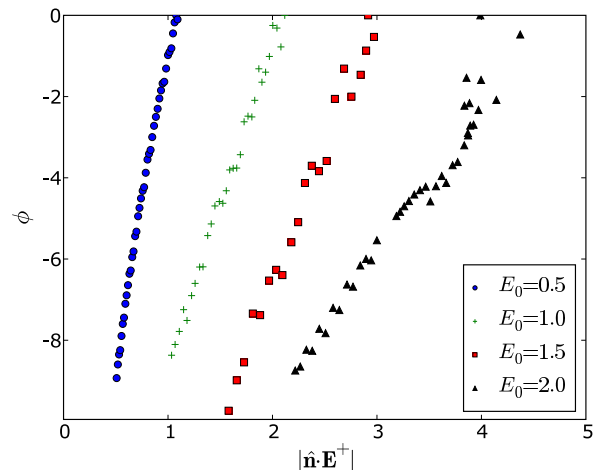}
\caption{$\varphi^+$ as a function of $E^+$ along the interface for different background fields $E_0$. A
gauge constant $\varphi^+(E_{max}^+)$ is subtracted from $\varphi^+$ for each data set.  For each electric
field, the data was extracted at times long enough for the space charge layer to be fully developed, but
always before the streamer branches.  These times were $t = 500$ for $E_0 = 0.5$, $t = 120$ for $E_0 = 1.0$,
$t = 43.6$ for $E_0 = 1.5$ and $t = 35$ for $E_0 = 2.0$.} \label{fig06}
\end{figure}

\begin{figure}[!hbtp]
\includegraphics[width=0.5\textwidth]{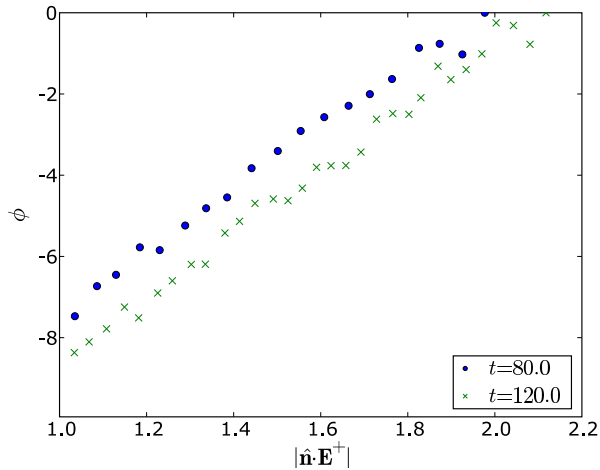}
\caption{$\varphi^+$ as a function of $E^+$ in a background field $E_0=1$ for times $t=80$ and $120$. The
slope is the same.} \label{fig07}
\end{figure}

The essential test of the boundary condition (\ref{GLsau}) is shown
in Fig.~\ref{fig06}. It shows how $\varphi^+$ depends on $|\hat{{\bf n}}\cdot{\bf E}^+|$ along the boundary
for several values of $E_0$ measured at times where the streamer is fully developed. For each set of data we
first determine $\varphi^+$ at the maximum of $E^+$, i.e. at $y=0$. This constant $\varphi^+(E_{max}^+)$ is
subtracted from all values $\varphi^+$ along the interface. Except for the smallest external field $E_0=0.5$,
the plots in Fig.~\ref{fig06} clearly are linear within the scatter of the data. Even for $E_0=0.5$ the
curvature is very small. (We note that with increasing $E_0$, the width $\ell_D$ of the diffusion layer decreases
and approaches the limiting spatial resolution of the numerics~\cite{CaroJCP}.
This explains the increasing scatter of the
data with increasing $E_0$). Furthermore, as is illustrated in Fig.~\ref{fig07} for $E_0 = 1.0$, for a fixed
$E_0$ the slope of the relation between $\varphi^+$ and $|\hat{{\bf n}}\cdot{\bf E}^+|$ does not depend on
time. Thus these numerical results can be summarized by the relation
\begin {equation} \label{GLwolf}
\varphi^+ = \varphi_0(E_0, t) - \ell(E_0)\, \hat{{\bf n}}\cdot{\bf E}^+ .
\end{equation}
Of course, neither the PDE-model nor the moving boundary approximation depend on the gauge $\varphi_0(E_0,t)$
which thus can be ignored. The prefactor $\ell(E_0)$ can be absorbed into the length scale of the moving
boundary model, with a compensating change of the time scale to preserve Eq.~(\ref{GLochse}). As mentioned
above, this rescaling also can absorb the enhancement of $v_n$ due to diffusion. As a result, the model
(\ref{GLratte})-(\ref{GLsau}) adequately appears to describe also fully developed curved streamer fronts.

We finally note that the parameter $\ell(E_0)$ decreases with increasing $E_0$, and it is well conceivable
that for $E_0\rightarrow\infty$ it tends to $1$, as predicted by our analysis of planar fronts. Furthermore
this behavior parallels the behavior of the thickness of the screening layer, suggesting the very plausible
assumption that it is this thickness which sets the spatial scale of the model also away from the limit
$E_0\rightarrow\infty$.


\section{Conclusions}
\label{conclusions}

Starting from a PDE-model of an anode-directed streamer ionization front, we have derived a boundary
condition valid for a moving boundary model of the streamer stage of the discharge. Due to the finite width
of the space charge layer surrounding the streamer head, in a moving boundary approximation the electric
potential has to be discontinuous across the boundary, and the boundary condition (\ref{GLsau}) proposed here
accounts for this jump in a very simple way. Our analytical derivation is restricted to planar fronts in
extreme external fields $E_0\rightarrow\infty$, but the analysis of numerical solutions of the PDE-model
shows that the boundary condition also applies to (two-dimensional) curved ionization fronts in weaker
external fields. We conclude that the moving boundary model adequately represents the evolution of negative
streamer fronts. This conclusion can also be drawn from studies of periodic arrays of interacting streamers
that show strong similarities with Saffman-Taylor fingers and will be presented elsewhere~\cite{R11}.

As with other moving boundary models in two dimensions, we now are in a position to use powerful conformal
mapping techniques to analytically attack questions like the stability of streamers against branching. Some
first results can be found in Refs.~\cite{PRL05,SIAM06}.

The moving boundary model does not explicitly refer to the interior of the streamer and thus leaves open
questions concerning the role of the residual electric field and the resulting currents inside the streamer.
Analyzing such questions within the framework of the minimal PDE model should lead to a more detailed
understanding of the structure of the space charge layer for curved fronts and should clarify the physics
underlying the phenomenological length parameter $\ell(E_0)$ occurring in Eq.~(\ref{GLwolf}). This problem,
which is important for fully understanding the physics of the streamer, is left for future work.
\\

{\bf Acknowledgement:} F.B.\ and A.L.\ were both supported by the Netherlands Organization for Scientific
Research NWO, F.B.\ through project 633.000.401 within the program "Dynamics of Patterns", and A.L.\ by the
Foundation for Technological Sciences STW, project CTF.6501. B.M.\ acknowledges a Ph.D.\ grant from CWI.


\appendix

\section{Bounds on $\varphi^+ -\varphi^-$}

Basic for our discussions are the properties of $\alpha(\eta)$ quoted in Eq.~(\ref{alphagen}), valid for
$\eta\ge 0$:
\begin{eqnarray}
    &&0\le \alpha(\eta)\le 1,\quad \lim_{\eta\rightarrow\infty} \alpha(\eta)=1, \label{eqA1} \\
    &&\frac{d\alpha}{d\eta}\ge 0. \label{eqA2}
\end{eqnarray}
We furthermore add the physically reasonable condition that
\begin{equation}
    \label{eqA3}
    0<b=\sup_{\eta\ge 0}\left(-\frac{d\alpha(1/\eta)}{d\eta}\right)<\infty
\end{equation}
exists, so that $\alpha(\eta)$ obeys the bound
\begin{equation}
    \label{eqA4}
    \alpha(\eta)\ge 1 - \frac{b}{\eta}.
\end{equation}
We now rewrite Eq.~(\ref{rho}) for $\rho(E)$ as
\begin{equation}
    \label{eqA5}     \rho(E)=(|E^+|-|E|)\;\bar\rho\left(\frac {E}{E^+},|E^+|\right),
\end{equation}
where
\begin{equation}
    \label{eqA6}
    \bar\rho(z,|E^+|)=\int_0^1 \alpha\left(|E^+|\left(z+(1-z)y\right)\right)\, dy.
\end{equation}
Eq.~(\ref{rp6}) for $\varphi^+ - \varphi^-$ takes the form
\begin{equation}
    \label{eqA7}
    \frac{\varphi^+-\varphi^-}{|E^+|}=\int_0^1 \frac{dx}{\bar\rho(x,|E^+|)}.
\end{equation}

The assumption (\ref{eqA1}) that $\alpha(E)\le 1$ for all $E$, leads directly to the lower bound
\begin{equation}
    \label{eqA8}
    \frac{\varphi^+-\varphi^-}{|E^+|}\ge 1.
\end{equation}
We note that a better lower bound can be obtain from the fact that since
$\alpha\left(E^+\left(x(1-y)+y\right)\right)$ increases with $x$, the function $\bar\rho(x,E^+)$ obeys
\begin{displaymath}
\bar\rho(x,|E^+|)\le \alpha(|E^+|).
\end{displaymath}
This leads to the improved lower bound
\begin{equation}
    \label{eqA9}
    \varphi^+-\varphi^-\ge \frac{|E^+|}{\alpha(|E^+|)},
\end{equation}
illustrating that weak fields cannot be screened since $|E^+|/\alpha(|E^+|)$ typically diverges for
$|E^+|\rightarrow 0$.

To derive an upper bound valid for large fields, we assume $|E^+|>b$ and split the integral in
Eq.~(\ref{eqA7}) as
\begin{equation}
    \label{eqA10}
    \int_0^1 \frac{dx}{\bar\rho(x,|E^+|)}=I_1+I_2,
\end{equation}
where
\begin{eqnarray}
    \label{eqA11}
    I_1&=&\int_0^{b/|E^+|}\frac{dx}{\bar\rho(x,|E^+|)}, \\
    I_2&=&\int_{b/|E^+|}^1\frac{dx}{\bar\rho(x,|E^+|)}.
\end{eqnarray}
By virtue of Eq.~(\ref{eqA2}), $\bar\rho(x,|E^+|)$ increases with $x$, which immediately yields the bound
\begin{displaymath}
    I_1\le \frac{b}{|E^+|} \bar\rho(0,|E^+|)^{-1}.
\end{displaymath}
Evaluating $\bar\rho(0,|E^+|)$ with the bound (\ref{eqA4}) on $\alpha(\eta)$ yields
\begin{equation}
    \label{eqA12}
    I_1\le \frac{b}{|E^+|}\left[ 1- \frac{b}{|E^+|}\left(1+\ln(|E^+|/b)\right)\right]^{-1}.
\end{equation}
To evaluate $I_2$ we write
\begin{eqnarray}
\bar\rho(x,|E^+|)&=&\int_0^x \alpha(|E^+|(x+(1-x)y))\, dy \nonumber \\
                 &+&\int_x^1 \alpha(|E^+|(y+(1-y)x))\, dy \nonumber \\
                 &>&\int_0^x \alpha(|E^+| x)\, dy+\int_x^1 \alpha(|E^+|y)\, dy, \nonumber \\
                 &>&1-\frac{b}{|E^+|}+\frac{b}{|E^+|}\ln x. \nonumber
\end{eqnarray}
This result yields
\begin{eqnarray}
    I_2&<&1-\frac{b/|E^+|}{1-b/|E^+|}\int_{b/|E^+|}^1\frac{\ln x}{1-b(1-\ln x)/|E^+|}\, dx, \nonumber \\
       &<&1-\frac{b/|E^+|}{1-b/|E^+|} \times \nonumber \\
       &\times& \frac{1}{1-b(1-\ln(b/|E^+|))/|E^+|}\int_{b/|E^+|}^1 \ln x\, dx, \nonumber \\
       &=& \frac{1}{1-b/|E^+|} \label{eqA13}.
\end{eqnarray}
Collecting all the results (and recalling $b/|E^+|<1$), we found in this appendix that
\begin{equation}
    \label{eqA14}
    \frac{\varphi^+-\varphi^-}{|E^+|}<\frac{1}{1-b/|E^+|}+\frac{b/|E^+|}{1- \frac{b}{|E^+|}
    \left(1+\ln(|E^+|/b)\right)},
\end{equation}
which for large $|E^+|/b$ leads to the bound (\ref{GLadler}) given in the main text. We note, in particular,
that $\varphi^+-\varphi^-$ does not contain a contribution of order $\ln(|E^+|/b)$, so that the leading
(constant), correction to $\varphi^+-\varphi^-=-E^+$ can be gauged away.

We finally note that Eqs.~(\ref{eqA6}), (\ref{eqA2}) imply that $\bar\rho(x,|E^+|)$ increases monotonically
with $|E^+|$, and thus that $(\varphi^+-\varphi^-)/|E^+|$ decreases monotonically.


\end{document}